\documentclass[10pt,conference,letterpaper]{IEEEtran}

\usepackage[T1]{fontenc}
\usepackage{amsmath,amssymb,amsfonts}
\usepackage{graphicx}
\usepackage{tikz}
\usepackage{xcolor}
\usepackage{multirow}
\usepackage{array}
\usepackage{booktabs}
\usepackage{textcomp}
\usepackage[hyphens]{url}
\usepackage{cite}
\usepackage[hidelinks]{hyperref}

\hypersetup{
  pdftitle={On the Limits of Machine-Learned Ranking for Modern Microarchitectural Policies},
  pdfauthor={Yanxin Zhang; Shayne Wadle; Yuxuan Xiong; Zheyu Fu; Trivikram Krishnamurthy; Karthikeyan Sankaralingam},
  pdfsubject={Microarchitectural performance prediction and policy ranking},
  pdfkeywords={computer architecture, machine learning, performance prediction, design-space exploration},
  pdflang={en-US}
}

\title{On the Limits of Machine-Learned Ranking for Modern Microarchitectural Policies}

\author{
  \IEEEauthorblockN{
    Yanxin Zhang\textsuperscript{$\ast\dagger$},
    Shayne Wadle\textsuperscript{$\dagger$},
    Yuxuan Xiong\textsuperscript{$\ast$},
    Zheyu Fu\textsuperscript{$\ast$},\\
    Trivikram Krishnamurthy\textsuperscript{$\ast$},
    Karthikeyan Sankaralingam\textsuperscript{$\ast\dagger$}
  }
  \IEEEauthorblockA{
    \textsuperscript{$\ast$}NVIDIA \qquad
    \textsuperscript{$\dagger$}University of Wisconsin--Madison\\
    \href{mailto:yanxinz@nvidia.com}{\nolinkurl{yanxinz@nvidia.com}},
    \href{mailto:swadle@cs.wisc.edu}{\nolinkurl{swadle@cs.wisc.edu}},
    \href{mailto:bearx@nvidia.com}{\nolinkurl{bearx@nvidia.com}},\\
    \href{mailto:zheyuf@nvidia.com}{\nolinkurl{zheyuf@nvidia.com}},
    \href{mailto:trivikramk@nvidia.com}{\nolinkurl{trivikramk@nvidia.com}},
    \href{mailto:karu@cs.wisc.edu}{\nolinkurl{karu@cs.wisc.edu}}
  }
}

\begin{document}
\maketitle
\thispagestyle{plain}
\pagestyle{plain}

\begin{abstract}
Machine-learning predictors estimate processor performance far faster than
cycle-level simulation. For design-space exploration, however, the valuable
test is not merely reproducing the usual hardware ordering, but identifying
how different hardware configurations rank on individual program phases. We evaluate four ML-predictors in two
design regimes: \emph{Structural Parameters} (SP), varying hardware
resources such as issue width, ROB size, and cache capacity; and
\emph{Behavioral Policies} (BP), varying prefetching and replacement
algorithms.
In the SP regime, aggregate ranking is strong, yet counter-intuitive
windows (CIW)---where the configuration expected to be slower is faster---constitute
$22.4\%$ of non-tied windows across five pairs with a clear architectural
prior. CIW match across these pairs is only $23.3$--$39.9\%$; every point
estimate is below the $50\%$ random strict-ordering reference.
The BP regime presents a different failure: ground-truth ties cover $37.8\%$
of pair-windows, most strict pairs have margins of only a few cycles, and no
model family reliably beats a feature-free majority baseline. NeuroScalar and
SimNet fall below that baseline, Concorde is statistically tied with it, and
the best selected OneDSE head improves by only $2.1$ percentage points.
Accuracy rises mainly at large margins.
We further show that this failure is not a matter of model capacity:
an information-theoretic analysis reveals that when ranking outcomes
depend on hidden microarchitectural state absent from the instruction
stream, no trace-based predictor can exceed the Bayes accuracy
determined by observable inputs alone.
Thus high cycle or aggregate ranking accuracy can reflect mastery of
easy, high-margin cases while missing the local reversals that carry the
most architectural insight and for which cycle-level simulation remains
indispensable.
\end{abstract}

\section{Introduction}\label{sec:intro}

Cycle-level simulation remains the reference methodology for evaluating new
microarchitectural ideas, but its speed makes broad design-space exploration
(DSE) difficult.
This has motivated a growing body of machine-learning-based performance
predictors.
Ithemal predicts basic-block throughput from static
code~\cite{mendis2019ithemal}; SimNet and TAO use deep learning to
accelerate architectural simulation~\cite{10.1145/3530891,10.1145/3656012};
Concorde combines analytical and ML models for fast CPU performance
modeling~\cite{10.1145/3695053.3731037}; OneDSE uses Transformer-based
workload-aware prediction for CPU metric prediction and design-space
exploration~\cite{onedse2025}; and NeuroScalar proposes lightweight
in-the-wild prediction from microarchitecture-independent
traces~\cite{neuroscalar2026}.
These systems share an appealing promise: if a learned model can approximate
simulator outputs cheaply, and thus run fast, architects can evaluate many more workloads and
candidate designs.

This paper focuses on a more specific question central to DSE: can ML
predictors reliably \emph{rank} candidate designs, especially when the
result contradicts a reasonable hardware prior?
Architects rarely need only an aggregate cycle estimate; they need to know
whether design $A$ is better than design $B$, and by enough margin to
justify a hardware decision.
Once a direction is established---say, that a new prefetcher wins
overall---the immediate question is \emph{where} it wins: which phases
drive the gain, whether it regresses elsewhere, and whether the effect is
consistent or fragile.
This is precisely why architects invest in detailed cycle-level simulation
rather than relying only on coarse IPC averages.

A time-resolved view, broken into program phases or \emph{instruction
windows} (contiguous groups of hundreds of instructions), exposes which
regions are responsible for a design's advantage and guides further
refinement.
ML predictors are naturally aligned with this granularity and could support
both aggregate ranking and per-region introspection.
Prior work, however, evaluates them primarily with aggregate regression
accuracy rather than asking whether each window preserves \emph{design order}.
A model can score well by reproducing the usual hardware direction yet fail
where that direction reverses; it can also misrank policies separated by
only a few cycles.
Conversely, modest absolute error can remain useful when the window-level
ordering is reliable.

We argue that the answer depends strongly on the design regime.
In the \emph{Structural Parameters} (SP) regime---changes to issue width,
ROB size, LSQ capacity, or cache capacity---resource differences tend to
create large, repeated performance gaps visible across many instruction
windows.
We expect enough margin for a learned model to recover the common ordering.
That average may nevertheless hide \emph{counter-intuitive windows} (CIWs),
in which the configuration expected to be faster instead loses: the dominant
trend is learnable, but the local exceptions where detailed simulation
provides new information may not be.
The \emph{Behavioral Policies} (BP) regime is harder on both counts.
Prefetching and replacement policies are already heavily optimized; they
often produce identical retirement-cycle counts across many windows, and
their wins depend on hidden policy state---replacement history, prefetch
timeliness, cache pollution---not present in the instruction stream.
We expect that in this regime the ranking label is frequently tied,
low-margin, or only weakly determined by observable instruction features.

We study this boundary with four input-compatible predictor families: a
NeuroScalar-style LSTM sequence predictor, a SimNet-style MLP/CNN latency
predictor, a Concorde-style window-summary predictor, and a OneDSE-style
Transformer predictor.
We deliberately evaluate models by downstream ranking quality under a common
protocol rather than by each paper's native accuracy metric alone.
All policy experiments use 29 traces, a chronological 80/10/10 split,
1500-instruction input windows, and middle-500-instruction targets.
The behavioral policy space contains eight competitive combinations of data prefetching,
instruction prefetching, and replacement policies, plus a stride/stride/LRU
baseline.

The results confirm the two-regime picture.
In the SP regime, aggregate structural ranking is strong: $77$--$89\%$ all-window agreement. Yet, this masks a failure on the reversals that matter most.
CIWs account for $22.4\%$ of non-tied windows across the five clear-prior
pairs, and CIW match falls below the $50\%$ random-ordering reference for
all four families.
In the BP regime, the failure is different in character: $37.8\%$ of
pair-windows are ground-truth ties, most strict pairs have margins of only
a few cycles, and no model family reliably beats a feature-free
training-majority baseline.
The best selected OneDSE head improves over that baseline by only
$2.1\%$, with gains concentrated at large margins where the
comparison is already easy.

The central contribution of this paper is a limitation result, not another
claim that a particular predictor is universally accurate.
We show that ML-based microarchitectural prediction must be evaluated as a
ranking problem---with margins, tie rates, and pair baselines---not only by
regression error.
The CIW result adds a second requirement: high average accuracy can reflect
learning the dominant hardware trend without learning when that trend breaks.
More fundamentally, we show that this failure is not a matter of model
capacity or architecture: when ranking outcomes depend on hidden
microarchitectural state absent from the instruction stream, no trace-based
predictor can exceed the Bayes accuracy determined by observable inputs
alone, regardless of model expressiveness (\S\ref{sec:bayes-limit}).

This paper is organized as follows. \S\ref{sec:overview} formalizes the ranking problem and the two design
regimes.
\S\ref{sec:models} describes the four predictor families under test
and the rationale for the input-compatible instantiation.
\S\ref{sec:methodology} details the experimental protocol, datasets,
and ranking metrics.
\S\ref{sec:results} reports results across both regimes, including the
CIW diagnostic and per-benchmark BP analysis.
\S\ref{sec:analysis} explains the shared failure through margin
analysis and an information-theoretic Bayes limit on partial observability.
\S\ref{sec:discussion} identifies the conditions under which ML
prediction remains useful and discusses the dynamic-state tradeoff.
\S\ref{sec:related} surveys related work and
\S\ref{sec:conc} concludes.

\section{Problem Overview}\label{sec:overview}
\subsection{Prediction Is Not the Same as Ranking}

Most ML performance-prediction papers report regression-style metrics: mean error, exact rounded-cycle accuracy, or window accuracy. These are important, but they do not fully measure usefulness for microarchitectural design-space exploration. DSE is usually comparative. A predictor is useful only if it can preserve the ordering between candidate designs:
\begin{equation}
\mathrm{sign}(y_A-y_B) \approx \mathrm{sign}(\hat{y}_A-\hat{y}_B),
\end{equation}
where $y_A$ and $y_B$ are simulator-measured cycle counts and $\hat{y}_A$, $\hat{y}_B$ are model predictions.

This formulation reveals a simple but important limitation. If $|y_A-y_B|$ is large, a model can tolerate moderate regression error and still preserve the correct order. If $y_A=y_B$, strict ranking is undefined. If $|y_A-y_B|$ is tiny, any small prediction error can flip the order. Thus ranking depends on model quality but also on the \emph{margin distribution} of the design space.

Average ranking can still answer the wrong architectural question. For pairs
with a clear expected hardware ordering, predicting the usual advantage
mostly recovers a known resource prior. We separately evaluate the non-tied
windows in which the configuration expected to be slower is faster. These
CIWs measure whether the predictor recognizes \emph{when} the obvious ordering
fails.

\subsection{Two Design Regimes}

We evaluate two regimes. The first is the \emph{Structural Parameters} (SP)
regime with five out-of-order processor configurations. These configurations
vary resources such as width, LSQ entries, ROB size, register count, and
cache capacity. Such changes often create visible performance differences
across many instructions, even when the exact cycle count remains difficult
to predict. Their broad differences make all-window ordering relatively
easy, while five pairs with a clear architectural prior let us isolate
counter-intuitive reversals. The remaining pairs have no sufficiently clear
expected winner and are excluded from the CIW metric.

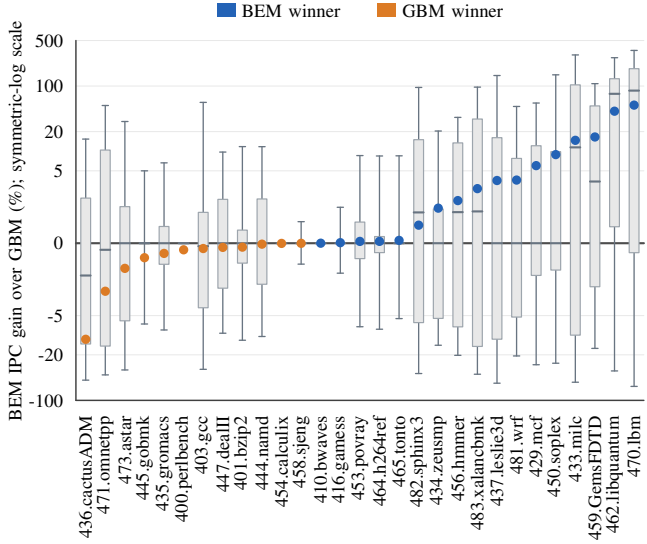
\begin{figure}[t]
    \centering
    \begin{tikzpicture}[x=0.102in,y=0.0046in,line cap=round,line join=round]
  \definecolor{bemdark}{RGB}{36,99,182}
  \definecolor{gbmdark}{RGB}{220,123,37}
  \definecolor{boxedge}{RGB}{104,116,129}

  \foreach \y/\lab in {
    0/{-100},51.643/{-20},96.125/{-5},178.219/{0},
    260.312/{5},304.795/{20},356.437/{100},408.080/{500}} {
    \draw[gray!18,line width=0.35pt] (-0.5,\y) -- (28.5,\y);
    \node[anchor=east,font=\scriptsize] at (-0.68,\y) {\lab};
  }
  \draw[black!75,line width=0.55pt] (-0.5,0) -- (-0.5,408.080);
  \draw[black!75,line width=0.55pt] (-0.5,0) -- (28.5,0);
  \draw[black!70,line width=0.8pt] (-0.5,178.219) -- (28.5,178.219);

  \foreach \i/\bench/\lo/\qlo/\med/\qhi/\hi/\point/\winner in {
    0/{436.cactusADM}/22.945/63.952/141.597/229.382/296.309/69.282/gbmdark,
    1/{471.omnetpp}/28.901/61.582/170.755/283.857/334.390/123.757/gbmdark,
    2/{473.astar}/34.505/90.234/178.219/219.698/316.239/149.716/gbmdark,
    3/{445.gobmk}/86.614/178.219/178.219/178.219/260.312/161.784/gbmdark,
    4/{435.gromacs}/79.845/154.337/178.219/197.200/269.349/166.662/gbmdark,
    5/{400.perlbench}/178.219/178.219/178.219/178.219/178.219/170.755/gbmdark,
    6/{403.gcc}/35.222/105.032/174.855/213.209/337.919/172.210/gbmdark,
    7/{447.dealII}/76.155/127.229/178.219/227.972/281.581/173.497/gbmdark,
    8/{401.bzip2}/68.194/155.650/178.219/193.077/287.768/173.780/gbmdark,
    9/{444.namd}/72.440/131.688/178.219/228.407/287.698/177.158/gbmdark,
    10/{454.calculix}/178.219/178.219/178.219/178.219/178.219/178.009/gbmdark,
    11/{458.sjeng}/154.509/178.219/178.219/178.219/202.651/178.106/gbmdark,
    12/{410.bwaves}/178.219/178.219/178.219/178.219/178.219/178.248/bemdark,
    13/{416.gamess}/144.249/178.219/178.219/178.219/218.974/178.830/bemdark,
    14/{453.povray}/83.514/160.767/178.219/202.187/277.607/180.277/bemdark,
    15/{464.h264ref}/80.692/167.566/178.219/185.777/277.152/180.320/bemdark,
    16/{465.tonto}/92.745/178.219/178.219/178.219/277.322/181.424/bemdark,
    17/{482.sphinx3}/30.440/88.189/213.152/295.619/354.872/198.689/bemdark,
    18/{434.zeusmp}/62.517/92.959/178.219/216.988/305.455/217.987/bemdark,
    19/{456.hmmer}/51.092/83.385/213.452/291.969/320.909/226.562/bemdark,
    20/{483.xalancbmk}/29.654/61.168/214.280/319.042/355.224/240.120/bemdark,
    21/{437.leslie3d}/19.487/69.377/178.219/297.892/368.388/249.335/bemdark,
    22/{481.wrf}/50.384/94.479/178.219/274.382/333.300/249.841/bemdark,
    23/{429.mcf}/40.425/141.693/178.219/288.745/337.188/266.202/bemdark,
    24/{450.soplex}/42.136/147.783/178.219/282.023/369.173/278.718/bemdark,
    25/{433.milc}/20.588/73.978/286.925/357.792/391.735/294.907/bemdark,
    26/{459.GemsFDTD}/58.952/128.873/248.185/333.929/359.155/298.688/bemdark,
    27/{462.libquantum}/33.395/196.903/347.636/364.743/388.591/327.928/bemdark,
    28/{470.lbm}/15.848/167.487/351.277/376.139/396.844/334.817/bemdark} {
    \draw[boxedge,line width=0.5pt] (\i,\lo) -- (\i,\qlo)
      (\i,\qhi) -- (\i,\hi);
    \draw[boxedge,line width=0.5pt] ({\i-0.16},\lo) -- ({\i+0.16},\lo)
      ({\i-0.16},\hi) -- ({\i+0.16},\hi);
    \filldraw[fill=gray!18,draw=boxedge!65,line width=0.45pt]
      ({\i-0.25},\qlo) rectangle ({\i+0.25},\qhi);
    \draw[boxedge,line width=0.8pt] ({\i-0.25},\med) -- ({\i+0.25},\med);
    \fill[\winner] (\i,\point) circle (1.75pt);
    \node[rotate=90,anchor=east,font=\scriptsize]
      at (\i,-7) {\bench};
  }

  \node[rotate=90,font=\scriptsize] at (-3.5,204.040)
    {BEM IPC gain over GBM (\%); symmetric-log scale};

  \node[anchor=south,font=\scriptsize] at (14,424) {%
    \textcolor{bemdark}{\rule{7pt}{5pt}}~BEM winner\qquad
    \textcolor{gbmdark}{\rule{7pt}{5pt}}~GBM winner};
\end{tikzpicture} 
    \caption{Signed per-window IPC gain of BEM
    (Berti--Entangling--Mockingjay) over GBM
    (Gaze--BARCA--Mockingjay) across 29 benchmarks. Boxes summarize the window
    distribution; colored markers identify each benchmark's aggregate winner.}
    \label{fig:selected-pair-ipc-difference}
\end{figure}

The second is the \emph{Behavioral Policies} (BP) regime with nine SOTA
policy configurations. These policies combine competitive prefetching and
replacement choices. Unlike structural changes, policy variants often affect
a small subset of instructions and frequently produce identical cycle counts.
Their differences may also depend on internal policy state that is not
directly present in the instruction stream. This regime stresses whether ML
predictors can distinguish highly optimized designs. Figure~\ref{fig:selected-pair-ipc-difference} illustrates this complexity
concretely: within a single benchmark, different program phases prefer
different policy combinations, and across benchmarks the aggregate winner
reverses---some applications favor one stack, others favor the other.
This per-phase volatility and cross-benchmark disagreement quantify why
BP ranking at instruction-window granularity is a fundamentally harder
task than aggregate IPC comparison (the next section outlines these policy configurations).

\subsection{Representative Predictor}

Our experiments use a sequence predictor as a representative ML-based
performance model. It consumes microarchitecture-independent instruction
features, including PC, opcode class, memory address, and register identifiers,
and predicts cycle counts for the center of a sliding instruction window. This
setup is intentionally favorable to in-the-wild deployment because it does not
require microarchitectural outcome features such as cache-hit level or
branch-predictor state. The same restriction also exposes the limitation we
study: if policy-specific ranking requires hidden microarchitectural state,
architecture-independent features may be insufficient.

\subsection{Key Takeaway}

The key question is not whether ML predictors can ever be useful. The
SP result demonstrates a rankable regime for broad structural effects. The
harder question is whether ML adds information where architectural intuition
is insufficient: SP counter-intuitive reversals and lead exchanges among
closely matched BP policy stacks. Across both regimes, the tested feature
set and supervision are least reliable on these decision-relevant windows.

\section{Predictor Families Under Test}\label{sec:models}

This paper does not propose another predictor architecture. Instead, it asks
whether the ML performance predictors that recent work has shown to
be accurate can support a stricter downstream task: recognizing
context-dependent order, including reversals of structural dominance and lead
changes among modern policies. To avoid tying the conclusion to one network,
we evaluate four families covering the main styles used by recent
ML-for-architecture work: NeuroScalar-style sequence prediction, SimNet-style
latency prediction, Concorde-style window-summary prediction, and OneDSE-style
Transformer prediction. Each model is trained and evaluated on aligned traces
and targets, then scored with the same ranking metrics.

The key methodological point is that we treat these models as
\textit{representatives of predictor families}, not as exact
reimplementations of every paper's full system. Each of these systems is motivated by the value of predicting performance
from lightweight, microarchitecture-independent traces---without requiring
simulator-only dynamic state such as cache-hit level or replacement metadata.
We instantiate each family in that deployment mode, which is the
lightweight setting each paper is primarily motivated by.
The question is whether this advertised lightweight deployment suffices for
the ranking task, not whether a richer simulator oracle could supply the
answer.
We train each family independently per configuration, giving every model
maximum specialization for its target and removing distributional mismatch
between configurations---strictly more favorable than any cross-architecture
training regime, so ranking failure cannot be attributed to insufficient
model capacity or training data.
For Concorde, whose full system pairs an analytical front-end (per-resource
throughput models from lightweight cache and branch simulation) with a
lightweight MLP, we instantiate only the ML backbone over instruction-stream
window statistics; reproducing the analytical stage would itself require
architecture-specific simulation outside the scope of a trace-based study,
and the result is a precise test of whether that backbone alone suffices
for ranking.

\subsection{Input Features}
For each retired instruction, the input trace records six
microarchitecture-independent fields: program counter, opcode class, memory
address, source register 1, source register 2, and destination register. These
are the same type of inputs that a lightweight trace-based predictor can
observe without exposing simulator-only state such as cache-hit level,
prefetch confidence, replacement metadata, MSHR occupancy, or
branch-predictor state. This restriction is central to the paper: if a
structural reversal or policy lead change depends primarily on hidden state,
an instruction-feature predictor may not have enough information to identify
it, regardless of whether it uses an LSTM, CNN/MLP, window-summary model, or
Transformer.

\subsection{Windowed Prediction Task}
All sequence-based models operate on instruction windows rather than isolated instructions. In the policy experiments, each sample uses 500 instructions of left context, 500 target instructions, and 500 instructions of right context. The sequence models receive the full 1500-instruction feature window and predict cycle counts for the middle 500 instructions. This matches the analysis granularity used in the ranking experiments: ordering is evaluated at the instruction/window level for the target region, not only as an aggregate benchmark-level IPC estimate.

Formally, for an input sequence $x_{1:N}$ and target length $R$, the target segment is centered in the window:
\begin{equation}
  x_{1:N}=\left[x_{1:s}\ \|\ x_{s+1:s+R}\ \|\ x_{s+R+1:N}\right],\quad
  \widehat{\mathbf y}=f_\theta(x_{1:N}),
\end{equation}
where $s=(N-R)/2$ and $\widehat{\mathbf y}\in\mathbb{R}^{R}$ contains predicted cycle counts for the middle instructions. In our main setting, $N=1500$ and $R=500$.

\subsection{Model-Specific Adaptations}
The LSTM emphasizes ordered context, the CNN/MLP local latency patterns, the
summary model low-variance distributional features, and the Transformer
long-range interaction. Testing all four separates a model-choice failure
from a limitation shared across inductive biases.

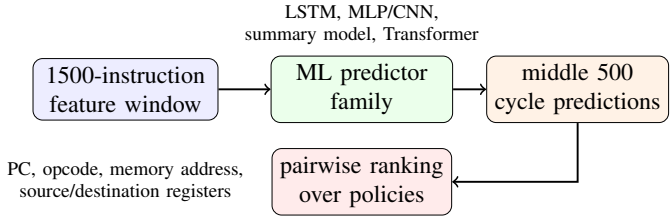
\begin{figure}[t]
  \centering
  \resizebox{\columnwidth}{!}{%
  \begin{tikzpicture}[x=1cm,y=1cm]
    \node[draw, rounded corners, align=center, minimum width=2.7cm, minimum height=0.72cm, fill=blue!7] (in) at (0,0) {1500-instruction\\feature window};
    \node[draw, rounded corners, align=center, minimum width=2.65cm, minimum height=0.72cm, fill=green!8] (models) at (3.45,0) {ML predictor\\family};
    \node[draw, rounded corners, align=center, minimum width=2.35cm, minimum height=0.72cm, fill=orange!10] (mid) at (6.6,0) {middle 500\\cycle predictions};
    \node[draw, rounded corners, align=center, minimum width=2.6cm, minimum height=0.72cm, fill=red!8] (rank) at (3.45,-1.35) {pairwise ranking\\over policies};
    \draw[->, thick] (in) -- (models);
    \draw[->, thick] (models) -- (mid);
    \draw[->, thick] (mid.south) |- (rank.east);
    \node[align=center, font=\footnotesize] at (0,-1.35) {PC, opcode, memory address,\\source/destination registers};
    \node[align=center, font=\footnotesize] at (3.45,0.95) {LSTM, MLP/CNN,\\summary model, Transformer};
  \end{tikzpicture}%
  }
  \caption{Common evaluation pipeline. Each model receives the same
  trace-derived features and produces cycle predictions; evaluation asks
  whether they preserve strict policy order and detect counter-intuitive
  structural reversals.}
  \label{fig:predictor-schematic}
\end{figure}

\subsection{Input-Compatible Reproduction Adaptations}
\label{sec:reproduction-adaptations}

Table~\ref{tab:reproduction-adaptations} states what is retained and what
changes when every family is placed on the common trace-observable
policy-ranking task.

\begin{table}[!b]
  \centering
  \scriptsize
  \caption{Model-family adaptations for the common experiment.}
  \label{tab:reproduction-adaptations}
  \setlength{\tabcolsep}{3pt}
  \renewcommand{\arraystretch}{0.9}
  \begin{tabular}{p{0.20\columnwidth}p{0.72\columnwidth}}
    \toprule
    Family & Retained mechanism and task-specific adaptation \\
    \midrule
    NeuroScalar & BiLSTM over 1500 instructions; export middle-500 cycles for strict pair ranking. \\
    SimNet & Direct CNN/MLP latency prediction, but without unavailable dynamic processor-state fields. \\
    Concorde & Distributional window summaries; predicts cycle totals without per-instruction regression outputs. \\
    OneDSE & Transformer sequence encoder; choose regression or classification head using validation only. \\
    \bottomrule
  \end{tabular}
\end{table}

\section{Experimental Methodology}\label{sec:methodology}

\newcommand{\base}[0]{8w}
\newcommand{\slow}[0]{6w+ls}
\newcommand{\ew}[0]{rob}
\newcommand{\mem}[0]{lsq}
\newcommand{\fourw}[0]{4w+mem}
\newcommand{\polbbm}{BBM}\newcommand{\polbbp}{BBP}\newcommand{\polbem}{BEM}\newcommand{\polbep}{BEP}
\newcommand{\polgbm}{GBM}\newcommand{\polgbp}{GBP}\newcommand{\polgem}{GEM}\newcommand{\polgep}{GEP}\newcommand{\polssl}{SSL}

\paragraph{Experimental question}
Recent ML performance predictors report strong accuracy for latency, CPI,
throughput, or design-space metrics. Our experiment asks a different question:
after a predictor is trained to estimate cycle behavior, can it recover the
ordering that an architect does not already know---local reversals of a valid
structural dominance prior and lead changes among closely matched policies?
We therefore evaluate models by downstream ranking quality, not only by
regression error or average ordering.

\paragraph{Trace collection and targets}
We use an instrumented ChampSim simulator~\cite{gober2022championship} to emit instruction features and ground-truth retirement-cycle counts. The predictor inputs are intentionally restricted to microarchitecture-independent features so that the experiment matches the deployment assumptions of lightweight trace-based predictors: the model sees the instruction stream, not simulator-only outcomes such as cache-hit level, prefetch usefulness, replacement metadata, or branch-predictor state. For each benchmark, the same input trace layout is shared across candidate configurations or policies, and each candidate has its own ground-truth cycle-count target. This alignment lets us compare predicted and true ordering at the same instruction/window positions.

\paragraph{Common policy protocol}
The final policy experiment uses 29 SPEC CPU2006 traces. Each model uses 1500-instruction input windows, predicts or scores the middle 500-instruction target region, and advances by a stride of 500 instructions. We use a chronological 80/10/10 train/validation/test split with a two-window guard band at split boundaries to avoid overlap between 1500-instruction input windows. Model selection, early stopping, and head selection use validation data only. The test set contains approximately 20K windows per trace and nine policies, giving about 20.9M policy-pair windows over all 29 traces.

\begin{table}[t]
  \centering
  \small
  \caption{Scale of the common nine-policy experiment. Counts are from the held-out test region; a one-window export convention changes the final digit for two baselines.}
  \label{tab:experiment-scale}
  \setlength{\tabcolsep}{4pt}
  \renewcommand{\arraystretch}{0.94}
  \begin{tabular}{lr}
    \toprule
    Quantity & Value \\
    \midrule
    Application traces & 29 \\
    Simulated instructions per trace & 100M \\
    Policy configurations / pairs & 9 / 36 \\
    Input / target / stride & 1500 / 500 / 500 instr. \\
    Chronological split & 80\% / 10\% / 10\% \\
    Test windows per trace & $\approx 19{,}998$ \\
    Nominal test pair-windows & 20.878M \\
    Strict / tied pair-windows & 12.995M / 7.883M \\
    \bottomrule
  \end{tabular}
\end{table}

\paragraph{Predictor families}
We reproduce four input-compatible predictor families: a NeuroScalar-style sequence predictor, a SimNet-style MLP/CNN latency predictor, a Concorde-style window-summary predictor, and a OneDSE-style Transformer sequence predictor. All models are trained on the same train/validation/test partition for a given benchmark and candidate. All models output either cycle predictions or pairwise scores that are converted into the same pairwise ranking labels. The OneDSE-style result reported as ``selected'' chooses between regression and classification heads using validation data only, then evaluates the frozen choice on test.

\paragraph{Runs and model selection}
The NeuroScalar-style and SimNet-style results use one completed training seed over all 29 traces and policies. Concorde-style and OneDSE-style results use three seeds; tables report their mean and sample standard deviation where seed-level values are available. The selected OneDSE head is not a test oracle: regression versus classification is chosen independently for each trace using validation performance, and that frozen choice is evaluated once on the test block. This distinction matters because selecting the better head after seeing test labels would overstate deployable accuracy.

\paragraph{Structural Parameters (SP) dataset}
The first design space is the SP experiment: every benchmark
is run under five processor configurations. The baseline is an 8-wide
out-of-order processor, and the four variants change structural resources such
as issue width, LS units, ROB size, LSQ size, and cache capacity.
Table~\ref{tab:proc_configs} summarizes these configurations. This dataset is
a positive control for all-window ranking because width and major-queue
changes create larger, more systematic effects than policy changes. It also
supports the harder CIW diagnostic: five pairs with a clear architectural
prior expose whether a model detects the windows in which those broad trends
reverse.

\begin{table}[t]
  \centering
  \caption{Microarchitectural parameters for the five structural processor configurations evaluated. The baseline is an 8-wide out-of-order processor. L2 cache is 8MB across the board.}
  \renewcommand{\arraystretch}{0.75}
  \label{tab:proc_configs}
  \begin{tabular}{lccccc}
    \toprule
    \multirow{2}{*}{\textbf{Config}}
     & \textbf{Base} & \textbf{6-wide+} & \textbf{Large} & \textbf{Large} & \textbf{More} \\
     & \textbf{8-wide}&  \textbf{LS Unit} & \textbf{ROB} & \textbf{LSQ} & \textbf{Memory} \\
     Abbr. & \textit{(\base)} & \textit{(\slow)} & \textit{(\ew)} & \textit{(\mem)} & \textit{(\fourw)} \\
    \midrule
    Width & 8 & 6 & 8 & 8 & 4 \\
    LS Units & 1 & 2 & 1 & 2 & 2 \\
    LSQ Entries & 32 & 32 & 32 & 64 & 32 \\
    Num Regs & 256 & 256 & 512 & 256 & 256 \\
    ROB Size & 192 & 192 & 384 & 192 & 192 \\
    \midrule
    L1D\$ Size & 64KB & 64KB & 64KB & 64KB & 128KB \\
    L1I\$ Size & 64KB & 64KB & 64KB & 64KB & 64KB \\
    \bottomrule
  \end{tabular}
\end{table}

\paragraph{Behavioral Policies (BP) dataset}
The second design space is the BP experiment, motivated by the SP results. We fix the broad structural configuration (Base 8-wide) and vary only the policy stack. Eight candidates combine Berti~\cite{navarro2022berti} and Gaze~\cite{chen2025gaze} data prefetching, BARCA~\cite{gratz2020barca} and Entangling~\cite{ros2021cost} instruction prefetching, and Mockingjay~\cite{shah2022effective} and PAC-IPv~\cite{mostofi2025light} replacement. We also include a stride/stride/LRU baseline, denoted SSL. Table~\ref{tab:policy-configs} lists the abbreviations used in the results. This dataset is intentionally difficult: the policies are competitive, their effects are sparse, and many instruction windows produce identical ground-truth cycle counts.

\begin{table}[t]
  \centering
  \small
  \caption{Nine BP configurations used for the BP-ranking experiment. SSL is a stride/stride/LRU baseline included with the eight competitive policy combinations.}
  \label{tab:policy-configs}
  \setlength{\tabcolsep}{4pt}
  \renewcommand{\arraystretch}{0.9}
  \begin{tabular}{llll}
    \toprule
    ID & Data prefetch & Inst. prefetch & Replacement \\
    \midrule
    \polbbm & Berti & BARCA & Mockingjay \\
    \polbbp & Berti & BARCA & PAC-IPv \\
    \polbem & Berti & Entangling & Mockingjay \\
    \polbep & Berti & Entangling & PAC-IPv \\
    \polgbm & Gaze & BARCA & Mockingjay \\
    \polgbp & Gaze & BARCA & PAC-IPv \\
    \polgem & Gaze & Entangling & Mockingjay \\
    \polgep & Gaze & Entangling & PAC-IPv \\
    \polssl & Stride & Stride & LRU \\
    \bottomrule
  \end{tabular}
\end{table}

\paragraph{Downstream ranking tasks}
For the five SP configurations, we rank processors for each retiring
instruction/window and compute pairwise ordering, full-rank match,
best-configuration match, and Kendall-style agreement. We additionally report
a CIW metric. A CIW is a non-tied window in which the configuration expected
to be slower is faster. Five pairs have a clear architectural prior:
large LSQ versus 4-wide+more-memory, 6-wide+LS versus
4-wide+more-memory, large ROB versus base 8-wide, large LSQ versus base
8-wide, and large LSQ versus 6-wide+LS. The other five pairs have no
sufficiently clear expected winner and receive no CIW label. CIW match
conditions on the reversal windows, with predicted ties counted as incorrect.

For the nine BP configurations, we compute pairwise ordering over all
$\binom{9}{2}=36$ pairs and separately report ground-truth ties. A tie has no
strict order to recover, so counting it as correct would inflate ranking.
Together, CIW match and non-tied policy match test whether models learn
context-dependent ordering rather than only a common direction.

\paragraph{Tie-aware ranking metrics}
For trace $t$, let $\mathcal{C}_t$ contain all test-window policy pairs with a strict ground-truth order. Its pairwise match is
\begin{equation}
  M_t=\frac{1}{|\mathcal{C}_t|}\sum_{(a,b,w)\in\mathcal{C}_t}
  \mathbb{1}[\widehat{S}_{ab,w}=S_{ab,w}].
\end{equation}
The headline macro result, $T^{-1}\sum_t M_t$, gives every application equal weight. We additionally report micro match over all strict pairs, a tie-neutral Kendall-like score that counts concordant minus discordant pairs, and aggregate best-policy overlap. Predicted ties remain in the strict-pair denominator and are not counted as correct.

\paragraph{Why majority, not only random}
Random strict ordering has expected accuracy $50\%$, but policy pairs have unequal win priors. For pair $(a,b)$, the train-pair-majority baseline predicts the more frequent strict direction in the training block for every test window. It uses policy identity but no instruction features. If $\pi_{ab}=P(S_{ab}=+1)$ in training, its expected in-distribution accuracy is
\begin{equation}
  A^{\mathrm{maj}}_{ab}=\max(\pi_{ab},1-\pi_{ab}).
\end{equation}
Gain over this baseline measures whether a model learns \emph{when} an ordering changes with program context, rather than merely which policy usually wins.

\paragraph{Statistical inference}
We compute paired confidence intervals by resampling the same 29 trace identities 100,000 times. For multi-seed models, seeds are first averaged within each trace, and the paired model-minus-majority difference is then bootstrapped across traces. This preserves workload pairing and prevents traces with many strict windows from dominating the uncertainty estimate. We treat trace variation, rather than millions of correlated windows, as the primary source of generalization uncertainty.

\section{Results}\label{sec:results}

The paper evaluates two design regimes introduced in Section~\ref{sec:overview}:
the \emph{Structural Parameters} (SP) regime, where five processor
configurations vary hardware resources such as issue width, ROB size, and
cache capacity, and the \emph{Behavioral Policies} (BP) regime, where nine
competitive policy stacks combine modern prefetching and replacement
algorithms.
The two regimes stress different properties of a ranking predictor and
require separate analysis; results for each appear in dedicated subsections
below.

We organize the BP results first, around increasingly specific
diagnostic questions.
We begin by asking whether any model family beats a feature-free majority
baseline on window-level pairwise ranking across 29 traces
(Section~\ref{sec:headline-policy-results}).
We then decompose \emph{why} the answer is limited: what the dataset provides in terms of ties and cycle margins
(Sections~\ref{sec:ties}--\ref{sec:margins}), and why high regression
accuracy does not transfer to ranking accuracy
(Section~\ref{sec:regression-vs-ranking}).
We next ask whether models succeed at coarser granularity: benchmark-level
aggregate ranking (Section~\ref{sec:aggregate}) and per-workload signal
detection (Section~\ref{sec:workload-dependent}).
We then turn to the SP regime for a sharper diagnostic: do models detect
counter-intuitive windows where an expected hardware advantage reverses
(Section~\ref{sec:ciw}), and does strong all-window SP accuracy give a
misleadingly favorable impression of that capability
(Section~\ref{sec:allwindow})?

\subsection{BP Ranking Accuracy}
\label{sec:headline-policy-results}

\textbf{Goal.}
We ask whether any ML predictor family can reliably distinguish
which of two behavioral policies produces fewer cycles at instruction-window
granularity across 29 benchmarks.
The baseline is a train-pair-majority rule---a feature-free classifier that
memorizes, per policy pair, which policy wins more often in training. This isolates instruction-window signal from static policy priors.
Table~\ref{tab:headline-models} reports the result.

\begin{table}[t]
  \centering
  \scriptsize
  \caption{Window-level policy-ranking accuracy across 29 traces. Match rates exclude ground-truth ties. Macro statistics average per-trace results and use paired trace bootstrap confidence intervals.}
  \label{tab:headline-models}
  \setlength{\tabcolsep}{1.5pt}
  \renewcommand{\arraystretch}{0.90}
  \begin{tabular}{@{}lccc@{}}
    \toprule
    Model & Match [95\% CI] & Maj. & $\Delta$ [95\% CI] \\
    \midrule
    LSTM & 52.98 [51.15,54.92] & 55.47 & -2.48 [-4.20,-1.06] \\
    SimNet & 53.73 [51.69,55.98] & 55.36 & -1.63 [-3.58,-0.06] \\
    Concorde & 55.58 [53.55,57.71] & 55.36 & +0.22 [-0.94,+1.35] \\
    OneDSE & 57.51 [55.11,60.24] & 55.42 & +2.09 [+0.52,+3.68] \\
    \bottomrule
\end{tabular}
\vspace{1pt}

\parbox{\columnwidth}{\centering\tiny Match and majority are percentages;
$\Delta$ is in percentage points.}
\end{table}

\begin{table}[t]
  \centering
  \scriptsize
  \caption{Instruction-level counterpart to Table~\ref{tab:headline-models}.}
  \label{tab:headline-models-instruction}
  \setlength{\tabcolsep}{1.5pt}
  \renewcommand{\arraystretch}{0.90}
  \begin{tabular}{@{}lccc@{}}
    \toprule
    Model & Match [95\% CI] & Maj. & $\Delta$ [95\% CI] \\
    \midrule
    LSTM & 52.92 [51.62,54.66] & 51.08 & +1.84 [+0.94,+3.00] \\
    SimNet & 52.83 [51.36,54.70] & 51.08 & +1.75 [+0.79,+2.96] \\
    Concorde & 51.18 [50.04,52.52] & 51.08 & +0.10 [-0.73,+0.85] \\
    OneDSE & 54.09 [52.35,56.46] & 51.08 & +3.01 [+1.77,+4.72] \\
    \bottomrule
\end{tabular}
\vspace{1pt}

\parbox{\columnwidth}{\centering\tiny Match and majority are percentages;
$\Delta$ is in percentage points.}
\end{table}

\textbf{Findings.}
Three of four families fall at or below this baseline.
The NeuroScalar-style LSTM ($-2.48\%$) and SimNet-style CNN/MLP
($-1.63\%$) are statistically below it.
The Concorde-style summary predictor sits at $+0.22\%$
(CI $[-0.94, +1.35]$), indistinguishable from the baseline.
The selected OneDSE-style Transformer head is the only model with a
positive trace-bootstrap interval at $+2.09\%$, yet its absolute macro
match is $57.5\%$, which is only $2.1\%$ above a majority prior
that uses no features whatsoever.
Table~\ref{tab:headline-models-instruction} shows instruction-level results
are similar; SimNet gains $1.75\%$ over its instruction-level majority,
but this signal does not materialize at window level.
Figure~\ref{fig:macro-margin}(a) makes the trajectory explicit: increasing
model complexity from local CNN features to long-range Transformer attention
moves the improvement interval from clearly negative to modestly
positive.

\begin{figure}[t]
  \centering
  \definecolor{rankgray}{HTML}{68717A}
  \definecolor{rankblue}{HTML}{3F6C8F}
  \definecolor{rankorange}{HTML}{C47B2B}
  \definecolor{rankgreen}{HTML}{2F806C}
  \definecolor{rankpurple}{HTML}{7655A5}
  \begin{tikzpicture}[x=1cm,y=1cm,font=\scriptsize,
                      line cap=round,line join=round]
    \path[use as bounding box] (0,-4.25) rectangle (8.30,4.12);

    \node[font=\footnotesize\bfseries] at (4.15,3.92)
      {(a) Improvement over train-pair-majority};
    \foreach \x/\lab in {3.24/-4,4.12/-2,5.00/0,5.88/{+2},6.76/{+4}} {
      \draw[gray!18] (\x,1.10) -- (\x,3.52);
      \node[below=1.5pt] at (\x,1.10) {\lab};
    }
    \draw[gray!68,line width=.8pt] (5.00,1.10) -- (5.00,3.52);
    \draw[gray!45] (2.80,1.10) -- (7.20,1.10);
    \node at (5.00,0.54) {improvement over majority (pp)};
    \foreach \y/\lab in {3.25/{NeuroScalar LSTM},2.65/{SimNet CNN/MLP},
                           2.05/{Concorde summary},1.45/{OneDSE selected}} {
      \node[anchor=east] at (2.55,\y) {\lab};
    }

    \draw[rankblue,line width=1.05pt] (3.15,3.25) -- (4.53,3.25);
    \draw[rankblue,line width=.75pt] (3.15,3.13) -- (3.15,3.37)
      (4.53,3.13) -- (4.53,3.37);
    \fill[rankblue] (3.91,3.25) circle (1.8pt);
    \node[rankblue,anchor=west] at (7.36,3.25) {$-2.48$};

    \draw[rankorange,line width=1.05pt] (3.42,2.65) -- (4.97,2.65);
    \draw[rankorange,line width=.75pt] (3.42,2.53) -- (3.42,2.77)
      (4.97,2.53) -- (4.97,2.77);
    \fill[rankorange] (4.28,2.65) circle (1.8pt);
    \node[rankorange,anchor=west] at (7.36,2.65) {$-1.63$};

    \draw[rankgreen,line width=1.05pt] (4.59,2.05) -- (5.59,2.05);
    \draw[rankgreen,line width=.75pt] (4.59,1.93) -- (4.59,2.17)
      (5.59,1.93) -- (5.59,2.17);
    \fill[rankgreen] (5.10,2.05) circle (1.8pt);
    \node[rankgreen,anchor=west] at (7.36,2.05) {$+0.22$};

    \draw[rankpurple,line width=1.05pt] (5.23,1.45) -- (6.62,1.45);
    \draw[rankpurple,line width=.75pt] (5.23,1.33) -- (5.23,1.57)
      (6.62,1.33) -- (6.62,1.57);
    \fill[rankpurple] (5.92,1.45) circle (1.8pt);
    \node[rankpurple,anchor=west] at (7.36,1.45) {$+2.09$};

    \node[font=\footnotesize\bfseries] at (4.15,-0.16)
      {(b) Match versus ground-truth margin};
    \begin{scope}[shift={(0.95,-3.02)},x=1.17cm,y=.125cm]
      \foreach \y/\lab in {0/50,5/55,10/60,15/65,20/70} {
        \draw[gray!18] (0,\y) -- (6,\y);
        \node[anchor=east,xshift=-2pt] at (0,\y) {\lab};
      }
      \draw[gray!45] (0,0) -- (6,0);
      \foreach \x/\lab in {0/1,1/2,2/{3--5},3/{6--10},
                             4/{11--25},5/{26--50},6/{$>$50}} {
        \node[below=2pt] at (\x,0) {\lab};
      }
      \draw[rankgray,line width=.65pt]
        plot[mark=*,mark size=1.25pt] coordinates
        {(0,1.10)(1,-0.01)(2,0.79)(3,1.69)(4,3.21)(5,5.03)(6,14.11)};
      \draw[rankblue,line width=.65pt]
        plot[mark=*,mark size=1.25pt] coordinates
        {(0,0.13)(1,0.01)(2,0.18)(3,0.61)(4,1.72)(5,2.02)(6,8.33)};
      \draw[rankorange,line width=.65pt]
        plot[mark=*,mark size=1.25pt] coordinates
        {(0,-0.09)(1,0.56)(2,0.39)(3,1.20)(4,1.62)(5,3.55)(6,9.86)};
      \draw[rankgreen,line width=.65pt]
        plot[mark=*,mark size=1.25pt] coordinates
        {(0,0.32)(1,0.38)(2,0.71)(3,1.60)(4,2.63)(5,4.70)(6,14.67)};
      \draw[rankpurple,line width=.75pt]
        plot[mark=*,mark size=1.35pt] coordinates
        {(0,1.46)(1,1.04)(2,1.83)(3,2.96)(4,4.64)(5,6.00)(6,17.70)};
    \end{scope}
    \node[anchor=south east] at (0.82,-0.49) {match (\%)};
    \node at (4.46,-3.62) {ground-truth margin (cycles)};

    \draw[rankgray,line width=.75pt] (0.30,-4.08) -- (0.63,-4.08);
    \node[anchor=west] at (0.69,-4.08) {Majority};
    \draw[rankblue,line width=.75pt] (1.78,-4.08) -- (2.11,-4.08);
    \node[anchor=west] at (2.17,-4.08) {LSTM};
    \draw[rankorange,line width=.75pt] (2.90,-4.08) -- (3.23,-4.08);
    \node[anchor=west] at (3.29,-4.08) {SimNet};
    \draw[rankgreen,line width=.75pt] (4.16,-4.08) -- (4.49,-4.08);
    \node[anchor=west] at (4.55,-4.08) {Concorde};
    \draw[rankpurple,line width=.75pt] (5.86,-4.08) -- (6.19,-4.08);
    \node[anchor=west] at (6.25,-4.08) {OneDSE};
  \end{tikzpicture}
  \vspace{-3pt}
  \caption{Policy-ranking difficulty: only selected OneDSE has a
  trace-bootstrap interval above train-pair-majority (a), and all models
  remain near chance at small ground-truth margins (b).}
  \label{fig:macro-margin}
\end{figure}

\textbf{Takeaway.}
\emph{The $2.1\%$ gain exists but too narrow to be actionable: as
Figure~\ref{fig:macro-margin}(b) shows, it is concentrated in high-margin
windows that are already easy to rank, not in the small-margin windows that impact
a design decision.
No predictor family achieves the accuracy needed to replace cycle-level
simulation for closely matched BP choices.}

\subsection{Ground-Truth Ties and Margin Structure}
\label{sec:ties}

\textbf{Goal.}
We ask how much useful ranking train-data the $20.9$M policy-pair windows
actually contain, by decomposing them by ground-truth label type and cycle
margin.

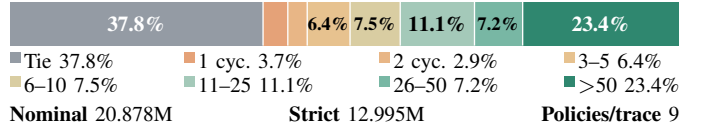
\begin{figure}[t]
  \centering
  \definecolor{pairtie}{HTML}{949BA4}
  \definecolor{pairone}{HTML}{E5A278}
  \definecolor{pairtwo}{HTML}{E8B686}
  \definecolor{pairthreefive}{HTML}{E7C28F}
  \definecolor{pairsixten}{HTML}{D9CDA2}
  \definecolor{paireleventwentyfive}{HTML}{A4C9B9}
  \definecolor{pairtwentysixfifty}{HTML}{78B7A4}
  \definecolor{pairoverfifty}{HTML}{2F876F}
  \begin{tikzpicture}[x=0.01\linewidth,y=1cm]
    \path[use as bounding box] (0,0) rectangle (100,0.58);
    \path[fill=pairtie]               (0,0)    rectangle (37.8,0.58);
    \path[fill=pairone]               (37.8,0) rectangle (41.5,0.58);
    \path[fill=pairtwo]               (41.5,0) rectangle (44.4,0.58);
    \path[fill=pairthreefive]         (44.4,0) rectangle (50.8,0.58);
    \path[fill=pairsixten]            (50.8,0) rectangle (58.3,0.58);
    \path[fill=paireleventwentyfive]  (58.3,0) rectangle (69.4,0.58);
    \path[fill=pairtwentysixfifty]    (69.4,0) rectangle (76.6,0.58);
    \path[fill=pairoverfifty]         (76.6,0) rectangle (100,0.58);
    \foreach \x in {37.8,41.5,44.4,50.8,58.3,69.4,76.6}
      \draw[white,line width=0.5pt] (\x,0) -- (\x,0.58);
    \node[font=\footnotesize\bfseries,text=white] at (18.9,0.29) {37.8\%};
    \node[font=\scriptsize\bfseries] at (47.6,0.29) {6.4\%};
    \node[font=\scriptsize\bfseries] at (54.55,0.29) {7.5\%};
    \node[font=\footnotesize\bfseries] at (63.85,0.29) {11.1\%};
    \node[font=\scriptsize\bfseries] at (73.0,0.29) {7.2\%};
    \node[font=\footnotesize\bfseries,text=white] at (88.3,0.29) {23.4\%};
  \end{tikzpicture}

  \vspace{2pt}
  {\footnotesize
  \setlength{\tabcolsep}{1pt}
  \begin{tabular*}{\linewidth}{@{\extracolsep{\fill}}llll@{}}
    \tikz[baseline=-0.5ex]\fill[pairtie] (0,0) rectangle (0.14,0.14);\,Tie 37.8\% &
    \tikz[baseline=-0.5ex]\fill[pairone] (0,0) rectangle (0.14,0.14);\,1 cyc. 3.7\% &
    \tikz[baseline=-0.5ex]\fill[pairtwo] (0,0) rectangle (0.14,0.14);\,2 cyc. 2.9\% &
    \tikz[baseline=-0.5ex]\fill[pairthreefive] (0,0) rectangle (0.14,0.14);\,3--5 6.4\% \\
    \tikz[baseline=-0.5ex]\fill[pairsixten] (0,0) rectangle (0.14,0.14);\,6--10 7.5\% &
    \tikz[baseline=-0.5ex]\fill[paireleventwentyfive] (0,0) rectangle (0.14,0.14);\,11--25 11.1\% &
    \tikz[baseline=-0.5ex]\fill[pairtwentysixfifty] (0,0) rectangle (0.14,0.14);\,26--50 7.2\% &
    \tikz[baseline=-0.5ex]\fill[pairoverfifty] (0,0) rectangle (0.14,0.14);\,$>$50 23.4\%
  \end{tabular*}

  \vspace{2pt}
  \begin{tabular*}{\linewidth}{@{\extracolsep{\fill}}ccc@{}}
    \textbf{Nominal} 20.878M &
    \textbf{Strict} 12.995M &
    \textbf{Policies/trace} 9
  \end{tabular*}}
  \caption{Composition of the 20.878M aligned policy-pair test windows. Ties provide no strict label; among strict pairs, $50.8\%$ have a margin of at most 25 cycles.}
  \label{fig:pair-composition}
\end{figure}

\begin{table}[t]
  \centering
  \scriptsize
  \caption{Micro-averaged non-tied pairwise match by pair group. Values are
  percentages; Concorde and OneDSE report mean $\pm$ sample standard
  deviation over three seeds.}
  \label{tab:pair-groups}
  \setlength{\tabcolsep}{2pt}
  \renewcommand{\arraystretch}{1.3}
  \begin{tabular}{@{}lrrrrrr@{}}
    \toprule
    Pair group & GT tie & Maj. & LSTM & SimNet & Conc. & OneDSE \\
    \midrule
    All & 37.76 & 56.82 & 53.77 & 54.62 & $56.84\!\pm\!0.52$ & $58.87\!\pm\!0.33$ \\
    Adv.--Adv. & 41.58 & 56.21 & 53.83 & 53.87 & $56.50\!\pm\!0.33$ & $58.08\!\pm\!0.38$ \\
    Adv.--SSL & 24.38 & 58.48 & 53.61 & 56.63 & $57.77\!\pm\!1.23$ & $61.00\!\pm\!0.21$ \\
    \bottomrule
  \end{tabular}
\end{table}

\textbf{Findings.}
Of all windows, $37.8\%$ are ground-truth ties carrying no strict
ranking label (Figure~\ref{fig:pair-composition}).
Of the remaining $12.99$M strict pairs, $50.8\%$ have a margin of at most
$25$ cycles.
We use the term \emph{Advanced (Adv)} to refer to the eight competitive policy
combinations from Table~\ref{tab:policy-configs} that pair modern SOTA prefetchers,  instruction prefetchers, and replacement policies; \emph{SSL} denotes the stride/stride/LRU baseline included as a
reference point.
The pair-group decomposition (Table~\ref{tab:pair-groups}) reveals
that Advanced--Advanced comparisons---the setting most relevant for choosing
among these SOTA proposals---show the weakest gains; the easier
Advanced--SSL distinction inflates combined numbers.

\textbf{Takeaway.}
\emph{Window counts overstate the ranking problem's tractability.
The data that actually distinguishes SOTA policies is sparse: a small
fraction of high-margin, non-tied, Advanced--Advanced comparisons carries
almost all the useful signal, and that fraction is small enough that even
strong models gain little.}

\subsection{Ranking Improves Mostly When the Margin Is Large}
\label{sec:margins}

\textbf{Goal.}
We ask a key realiability diagnostic: whether model gains are spread evenly across the margin distribution
or concentrated at specific cycle-difference thresholds. %
Figure~\ref{fig:macro-margin} groups non-tied windows by ground-truth
cycle margin.

\textbf{Findings.}
All models remain near chance at the smallest margin bins (1--5 cycles) and
absolute match rises most strongly in the $>50$-cycle bin, where policies are
no longer nearly indistinguishable; only Concorde and OneDSE meet or exceed
the majority baseline there.
This is what a margin-based view predicts: ranking is stable only
when the true policy difference is large relative to prediction error.
Figure~\ref{fig:macro-margin}(b) shows this pattern is consistent in all
four families.

\textbf{Takeaway.}
\emph{A predictor may achieve high instruction-level cycle accuracy because many
instructions are easy or tied, yet fail on narrow-margin windows where
policies differ by only one or two cycles---critical windows for design
decisions.}

\subsection{Regression Accuracy Does Not Imply Ranking Accuracy}
\label{sec:regression-vs-ranking}

\begin{table}[t]
  \centering
  \scriptsize
  \caption{Benchmark-level aggregate policy ranking. Cycles are summed over the test windows before ranking nine policies per trace; Concorde and OneDSE entries are seed means.}
  \label{tab:aggregate-ranking}
  \setlength{\tabcolsep}{2pt}
  \renewcommand{\arraystretch}{0.95}
  \resizebox{\columnwidth}{!}{%
    \begin{tabular}{lccc}
      \toprule
      Predictor & \shortstack{Macro match /\\Kendall-like} &
      \shortstack{GT-best\\overlap} & \shortstack{Exact best-set\\match} \\
      \midrule
      NeuroScalar-style LSTM & 54.64\% / 0.093 & 27.59\% & 27.59\% \\
      SimNet-style CNN/MLP & 57.39\% / 0.148 & 44.83\% & 17.24\% \\
      Concorde-style summary & 71.25\% / 0.425 & 44.83\% & 9.20\% \\
      OneDSE-style Transformer & 58.71\% / 0.174 & 31.03\% & 12.64\% \\
      \bottomrule
    \end{tabular}%
  }
\end{table}

\textbf{Goal.}
Prior ML performance-prediction work evaluates models on regression
metrics: mean error or tolerance-window accuracy.
We test whether high values on these metrics are sufficient evidence of
ranking quality.

\textbf{Findings.}
They are not.
The NeuroScalar-style LSTM achieves $95.8\%$ macro instruction-level
within-one-cycle accuracy, yet falls $2.48\%$ \emph{below} the
window-level majority baseline.
SimNet similarly achieves $95.6\%$ instruction-level within-one-cycle
accuracy while remaining below majority at window level.
Concorde shows the complementary failure: strong aggregate ranking in some
seeds, but two seeds exhibit catastrophic regression outliers on a single
trace, demonstrating that favorable ranking summaries can conceal regression
instability (Table~\ref{tab:aggregate-ranking}).

\textbf{Takeaway.}
\emph{Strong regression accuracy provides false confidence for DSE: a model that
looks best by standard metrics can sit below a feature-free
baseline on the decisions that matter.
This failure mode is invisible unless ranking is evaluated directly on
non-tied pairs broken down by margin.}

\subsection{Benchmark-Level Aggregate Ranking}
\label{sec:aggregate}

\textbf{Goal.}
We compute a coarser benchmark-level aggregate: predicted and ground-truth
cycles are summed over all test windows per trace, then all nine policies
are ranked once per benchmark. %
Crucially, errors on individual windows can cancel out in the summation that
makes this the most favorable possible test for aggregate model behavior.
If models cannot reliably identify the best policy even with this
error-cancellation advantage, the window-level failure is not a fluke of
granularity.

\textbf{Findings.}
Concorde-style achieves $71.25\%$ aggregate pairwise match with a
Kendall-like score of $0.425$. Its predicted-best set overlaps the
ground-truth best set in $44.83\%$ of cases but matches that set exactly in
only $9.20\%$. OneDSE-style Transformer achieves $58.71\%$ aggregate
pairwise match, a Kendall-like score of $0.174$, $31.03\%$ GT-best overlap,
and $12.64\%$ exact best-set match
(Table~\ref{tab:aggregate-ranking}).
Many benchmarks have tied ground-truth best policies, which reduces the data available for this task.

\textbf{Takeaway.}
\emph{The Concorde discrepancy makes this concrete: near-baseline at window level
yet $71\%$ Kendall overlap at aggregate.
A model can appear to rank policies well at benchmark granularity while
contributing nothing beyond the static training prior at the window
granularity where policy effects are actually phase-localized.}

\subsection{Signal Exists, but Is Workload-Dependent}
\label{sec:workload-dependent}

\begin{table}[t]
  \centering
  \caption{Per-trace strict pairwise accuracy (\%).
  Concorde and OneDSE entries are seed means.}
  \label{tab:trace-heterogeneity}
  \setlength{\tabcolsep}{1.5pt}
  \renewcommand{\arraystretch}{0.6}
  \resizebox{\columnwidth}{!}{%
    \begin{tabular}{lcccc}
      \toprule
      Trace & LSTM & SimNet & Concorde & OneDSE \\
      \midrule
      \texttt{400.perlbench} & 49.4 & 51.8 & 52.1 & 52.9 \\
      \texttt{401.bzip2} & 48.8 & 50.7 & 50.0 & 50.6 \\
      \texttt{403.gcc} & 57.4 & 56.4 & 62.1 & 63.1 \\
      \texttt{410.bwaves} & 52.1 & 51.2 & 51.2 & 53.9 \\
      \texttt{416.gamess} & 48.9 & 49.8 & 50.3 & 51.7 \\
      \texttt{429.mcf} & 57.9 & 60.9 & 62.5 & 62.9 \\
      \texttt{433.milc} & 62.7 & 47.5 & 57.3 & 53.5 \\
      \texttt{434.zeusmp} & 53.9 & 54.3 & 56.3 & 58.1 \\
      \texttt{435.gromacs} & 51.1 & 48.7 & 50.6 & 52.3 \\
      \texttt{436.cactusADM} & 54.4 & 56.7 & 57.9 & 57.9 \\
      \texttt{437.leslie3d} & 48.3 & 56.6 & 58.6 & 60.2 \\
      \texttt{444.namd} & 50.3 & 50.3 & 49.9 & 49.8 \\
      \texttt{445.gobmk} & 50.7 & 56.0 & 56.4 & 58.3 \\
      \texttt{447.dealII} & 49.4 & 50.5 & 49.7 & 50.0 \\
      \texttt{450.soplex} & 56.2 & 61.7 & 64.3 & 66.6 \\
      \texttt{453.povray} & 52.0 & 49.4 & 55.5 & 63.0 \\
      \texttt{454.calculix} & 49.9 & 50.1 & 50.2 & 50.2 \\
      \texttt{456.hmmer} & 50.6 & 50.6 & 53.3 & 53.4 \\
      \texttt{458.sjeng} & 50.3 & 48.9 & 51.8 & 52.9 \\
      \texttt{459.GemsFDTD} & 40.8 & 42.1 & 60.8 & 61.0 \\
      \texttt{462.libquantum} & 64.7 & 73.3 & 72.3 & 78.6 \\
      \texttt{464.h264ref} & 49.8 & 49.6 & 50.2 & 50.2 \\
      \texttt{465.tonto} & 54.9 & 54.9 & 56.1 & 58.7 \\
      \texttt{470.lbm} & 62.3 & 61.1 & 58.7 & 75.6 \\
      \texttt{471.omnetpp} & 56.1 & 55.5 & 58.2 & 57.7 \\
      \texttt{473.astar} & 52.0 & 53.9 & 55.3 & 54.8 \\
      \texttt{481.wrf} & 49.6 & 48.8 & 43.3 & 51.5 \\
      \texttt{482.sphinx3} & 49.9 & 54.6 & 56.6 & 57.7 \\
      \texttt{483.xalancbmk} & 60.8 & 62.0 & 60.2 & 61.0 \\
      \midrule
      \textbf{Macro mean} & \textbf{52.98} & \textbf{53.73} &
      \textbf{55.58} & \textbf{57.51} \\
      \bottomrule
    \end{tabular}%
  }
\end{table}

\textbf{Goal.}
We ask whether the BP ranking failure is uniform across all 29 benchmarks
or whether specific workloads are genuinely learnable.
Table~\ref{tab:trace-heterogeneity} breaks down results per benchmark;
each reported percentage is that benchmark's strict pairwise match rate.

\textbf{Findings.}
There is substantial heterogeneity.
The LSTM beats its per-benchmark majority baseline on 4 of 29 benchmarks;
SimNet on 11; Concorde on 14; the selected OneDSE head on 21.
Requiring a practically visible gain of at least $3\%$ reduces these counts
to 2, 3, 4, and 10 benchmarks, respectively.
The strongest OneDSE gains occur on \texttt{lbm} ($+13.0\%$),
\texttt{povray} ($+11.0\%$), \texttt{leslie3d} ($+7.7\%$),
\texttt{omnetpp} ($+7.2\%$), and \texttt{tonto} ($+7.2\%$). All benchmarks
with repeated memory phases whose observable address and opcode patterns
correlate with a consistent policy winner.
The same model loses $10.9\%$ on \texttt{milc}, and models frequently
disagree about which benchmarks are learnable.

\textbf{Takeaway.}
\emph{A favorable global mean is not a reliability guarantee for a new benchmark.
Whether a new application belongs to the learnable minority cannot be
determined without per-benchmark evaluation, and a deployable predictor must
know this in advance.}

\subsection{SP Regime: Counter-Intuitive Windows}
\label{sec:ciw}

\begin{table}[t]
  \centering
  \scriptsize
  \caption{Counter-intuitive-window results (\%) on the five structural pairs with
  a clear architectural prior. Prior order lists the inverted order. GT CI is the fraction of non-tied windows of this order; model columns report match conditioned on those
  windows. Predicted ties count as incorrect.}
  \label{tab:ciw-structural}
  \setlength{\tabcolsep}{1.7pt}
  \renewcommand{\arraystretch}{0.88}
  \begin{tabular}{@{}lccccc@{}}
    \toprule
    Prior order & GT CI & LSTM & SimNet & Conc. & OneDSE \\
    \midrule
    \mem\,$<$\,\fourw & 13.2 & 22.4 & 14.7 & 14.7 & 29.4 \\
    \slow\,$<$\,\fourw & 13.6 & 29.6 & 10.3 & 21.8 & 28.3 \\
    \ew\,$<$\,\base & 43.5 & 36.1 & 35.7 & 44.6 & 41.9 \\
    \mem\,$<$\,\base & 23.1 & 52.0 & 40.7 & 21.5 & 58.0 \\
    \mem\,$<$\,\slow & 18.9 & 35.7 & 15.4 & 23.1 & 42.0 \\
    \midrule
    Avg. & 22.4 & 35.2 & 23.3 & 25.2 & 39.9 \\
    \bottomrule
  \end{tabular}
\end{table}

\begin{table}[t]
  \centering
  \scriptsize
  \caption{All-window structural ordering on 23 traces. All evaluates the
  displayed $A\leq B$ cycle predicate; Strict excludes ground-truth ties but
  counts predicted ties as incorrect. Each model cell reports All/Strict.
  Nonzero is the fraction with $y_A\neq y_B$. Values are \%; Concorde and OneDSE are seed means.}
  \label{tab:pairwise-full}
  \setlength{\tabcolsep}{2.1pt}
  \renewcommand{\arraystretch}{0.9}
  \begin{tabular}{@{}lccccc@{}}
    \toprule
    Pair & LSTM & SimNet & Conc. & OneDSE & Nonzero \\
    \midrule
    \fourw$\leq$\base & 77.2/77.4 & 78.8/79.0 & 83.7/83.8 & 77.8/77.9 & 99.7 \\
    \fourw$\leq$\ew   & 77.5/77.7 & 78.9/79.1 & 83.8/83.9 & 78.5/78.7 & 99.7 \\
    \fourw$\leq$\mem  & 83.9/84.1 & 87.4/87.5 & 86.9/87.0 & 84.5/84.6 & 99.8 \\
    \fourw$\leq$\slow & 84.5/84.7 & 86.4/86.7 & 85.9/86.0 & 84.2/84.3 & 99.7 \\
    \base$\leq$\ew    & 56.6/58.6 & 59.4/62.2 & 62.9/63.5 & 59.9/61.2 & 92.8 \\
    \base$\leq$\mem   & 57.5/57.4 & 64.7/65.3 & 75.3/76.4 & 62.9/63.4 & 97.5 \\
    \base$\leq$\slow  & 52.1/51.8 & 56.2/55.9 & 72.1/72.3 & 55.1/54.9 & 98.7 \\
    \ew$\leq$\mem     & 53.4/52.9 & 58.7/59.1 & 71.1/72.0 & 61.1/61.5 & 97.6 \\
    \ew$\leq$\slow    & 56.9/56.6 & 61.5/61.3 & 72.1/72.3 & 59.4/59.3 & 98.9 \\
    \mem$\leq$\slow   & 74.8/74.8 & 80.5/80.5 & 81.7/81.2 & 74.3/74.3 & 99.0 \\
    \bottomrule
  \end{tabular}
\end{table}

Having established the limits of BP ranking, we now turn to the SP regime,
where structural resource differences create larger and more systematic
performance gaps.

\textbf{Goal.}
The key question is not whether models can follow the dominant hardware
direction that a resource table already implies, but whether they detect
the exceptions: windows where the expected architectural advantage reverses.
We define a \emph{counter-intuitive window} (CIW) as a non-tied window where
the expected-to-be-faster configuration is instead slower, and we evaluate
all four families specifically on these reversals across the five SP pairs
that have a clear architectural prior.
Aggregate SP accuracy cannot answer this question because it is dominated
by the easy majority direction; CIW match isolates only the hard cases.

\textbf{Findings.}
Averaged across the five clear-prior pairs, CIWs account for $22.4\%$ of
non-tied windows.
All four model families fall \emph{below} the $50\%$ random strict-ordering
reference on these windows (Table~\ref{tab:ciw-structural}). %
No model reliably detects the windows where an expected architectural
advantage reverses.

\textbf{Takeaway.}
\emph{The models recover the dominant hardware trend but not the exceptions,
CIWs. In CIWs, detailed simulation provides information that
architectural intuition cannot, and aggregate accuracy does not establish
that a predictor can substitute for simulation on those decisions.}

\subsection{SP Regime: Why All-Window Accuracy Looks Strong}
\label{sec:allwindow}

\textbf{Goal.}
We now ask: is high aggregate SP accuracy meaningful evidence that models
handle the SP design space reliably?
Table~\ref{tab:pairwise-full} reports all-window ordering across all
processor configurations in the SP regime.

\textbf{Findings.}
The all-window task is substantially easier, particularly for pairs involving
the 4-wide configuration, where strict agreement reaches $77$--$88\%$
(Table~\ref{tab:pairwise-full}).
These high numbers reflect the large, systematic differences that width and
queue-resource changes create across most instruction windows---differences
that any calibrated predictor will learn by following the dominant direction.
The same predictors that score well here fall below chance on CIWs because
the all-window task does not require detecting local exceptions.

\textbf{Takeaway.}
\emph{ML is genuinely useful for broad SP pruning: models can eliminate
clearly inferior structural regions where margins are large and systematic.
The CIW boundary marks where that pruning should stop and direct simulation
should take over. A predictor that exceeds $80\%$ all-window accuracy can
simultaneously fall below $40\%$ on the reversals an architect most needs
to detect.}

\section{Why Modern Policy Ranking Is Hard}\label{sec:analysis}

The empirical results separate trend recovery from exception detection.
In the SP regime, aggregate structural ranking reaches approximately $91\%$
agreement, yet all four CIW point estimates fall below the $50\%$ random
strict-ordering reference for the five clear-prior pairs.
In the BP regime, ML predictors generally fail to beat a simple
feature-free baseline, demonstrating that the instruction stream carries
insufficient signal for reliable policy ranking.
This section formalizes the shared limits behind both results using tools
from statistical learning theory and information theory.
Section~\ref{sec:label-dist} shows that the nominal dataset size
substantially overstates the useful ranking supervision, due to ties and
temporal autocorrelation.
Section~\ref{sec:margin-amplifies} derives why even modest prediction error
can flip a ranking when the true performance margin is narrow and why this makes
regression accuracy an unreliable proxy for ranking quality.
Section~\ref{sec:bayes-limit} establishes a hard information-theoretic
ceiling: because the policy-critical hidden state ($H$) is absent from the
instruction trace, no model can exceed the Bayes accuracy determined by
$I(S;X)$ regardless of architecture or
data volume.
Sections~\ref{sec:tsne} and~\ref{sec:expressiveness} provide qualitative
feature-space evidence and reconcile why model expressiveness helps on some
benchmarks but not in general.

\subsection{Label Distribution and Effective Supervision}
\label{sec:label-dist}

Suppose a benchmark contributes $N$ windows and $K$ policies.
The nominal number of pair examples is $N\binom{K}{2}$.
If pair $(a,b)$ ties with probability $p^{(0)}_{ab}$, the expected number
of strict labels is only
\begin{equation}
  N^{\mathrm{strict}}_{ab}=N(1-p^{(0)}_{ab}).
  \label{eq:strict-samples}
\end{equation}
In our 500-instruction test windows, $37.8\%$ of nominal pairs are tied;
in the instruction-level diagnostic the tie rate reaches $99.15\%$.
A dataset with millions of nominal comparisons can therefore contain far
fewer useful strict labels.

Even Equation~\eqref{eq:strict-samples} overstates independent supervision.
Adjacent 1500-instruction inputs overlap in context and policy outcomes
persist through program phases.
If $\rho_k$ is the lag-$k$ autocorrelation of a pair label,
\begin{equation}
  N^{\mathrm{eff}}_{ab}\approx
  \frac{N^{\mathrm{strict}}_{ab}}
       {1+2\sum_{k\geq1}\rho_k},
  \label{eq:effective-samples}
\end{equation}
so 100 million retired instructions do not imply 100 million independent
examples of policy differentiation.

Label imbalance further separates apparent accuracy from learned signal.
Let $\pi_{ab}=P(S_{ab}=+1)$ on strict training examples.
A feature-free classifier obtains
$A^{\mathrm{maj}}_{ab}=\max(\pi_{ab},1-\pi_{ab})$.
This $55.42\%$ result shows that policy-pair priors already explain
a substantial portion of performance above random; gains must be measured
over $A^{\mathrm{maj}}$, not only over $50\%$.

The relevant sample size for training is therefore the number of non-tied
\emph{transitions among distinguishable policy regimes}---a quantity far
smaller than the nominal window count, which explains why even large BP
datasets yield weak ranking signal.

\noindent\textbf{Takeaway.} \emph{A dataset of 20M windows sounds large,
but most windows look identical to both policies (ties), neighboring windows
repeat the same phase (correlation), and some policy pairs have a
predictable winner without even looking at instructions (label imbalance).
After accounting for all three, the amount of genuinely informative data is
a small fraction of the headline number.}

\subsection{Ranking Margin Amplifies Regression Error}
\label{sec:margin-amplifies}

For pair $(a,b)$, the model predicts
\begin{equation}
    \widehat{\Delta}_{ab}=\Delta_{ab}+D_{ab}, \qquad
  D_{ab}=\epsilon_a-\epsilon_b.
\end{equation}
A rank error occurs when $D_{ab}$ crosses $-\Delta_{ab}$.
If the differential error is approximately zero-mean Gaussian with standard
deviation $\sigma_D$, the conditional error probability is
\begin{equation}
  P(\mathrm{error}\mid\Delta_{ab})
  =\Phi\!\left(-\frac{|\Delta_{ab}|}{\sigma_D}\right),
  \label{eq:rank-error}
\end{equation}
where $\Phi$ is the standard normal CDF.
For any error distribution a distribution-free bound follows: for any
threshold $\gamma>0$,
\begin{equation}
  P(\mathrm{misrank})\leq P(|\Delta|\leq\gamma)+P(|D|\geq\gamma).
  \label{eq:margin-error-bound}
\end{equation}
Equation~\ref{eq:margin-error-bound} exposes two independent levers.
More training can shrink the error tail $P(|D|\geq\gamma)$, but it cannot
remove probability mass from the design space's near-zero margin region
$P(|\Delta|\leq\gamma)$.
Choosing more widely separated candidates improves rankability even with
the same predictor.
Figure~\ref{fig:macro-margin}(b) follows this decomposition empirically:
every family approaches chance in the small-margin bins and improves only
as $|\Delta|$ grows.

Equation~\ref{eq:rank-error} also explains why better cycle regression may
have little effect on BP ranking.
Modern policies place much of their probability mass at $\Delta=0$ and
much of the remainder near zero.
A reduction in mean absolute error can improve the reported cycle metric
without moving enough samples into the stable-margin regime.
The SP regime occupies a different part of this distribution: width and
capacity changes affect many instructions, giving their dominant direction
a large margin-to-error ratio.
Conditioning on CIWs removes those easy cases and exposes the local
reversals; the resulting $23.3$--$39.9\%$ CIW match confirms that high
all-window agreement does not transfer to the tail where simulation is
most informative.

\noindent\textbf{In plain terms:}\emph{Getting a cycle count wrong by one cycle
usually does not matter---unless the two policies you are comparing differ
by exactly one cycle, in which case that same error flips your ranking.
BP policies often differ by just a handful of cycles per window.
This means a predictor that looks highly accurate on standard metrics can
still pick the wrong policy the majority of the time on the windows that
actually matter for a design decision.}

\subsection{Partial Observability Sets a Bayes Limit}
\label{sec:bayes-limit}

Let $X$ be the observed instruction window and $H$ the hidden
microarchitectural state: cache contents, outstanding misses, prefetch
queues and confidence, replacement history, and other policy metadata.
A strict pair label is generated by
\begin{equation}
  S_{ab}=g_{ab}(X,H).
\end{equation}

\begin{figure}[t]
  \centering
  \begin{tikzpicture}[x=0.92cm,y=0.92cm, every node/.style={font=\scriptsize}]
    \node[draw, rounded corners, align=center, fill=blue!7, minimum width=1.65cm]
      (x) at (0,1.2) {observed $X$\\PC/opcode/address};
    \node[draw, rounded corners, align=center, fill=green!8, minimum width=1.55cm]
      (z) at (3.0,1.2) {representation\\$Z=f_\theta(X)$};
    \node[draw, rounded corners, align=center, fill=orange!10, minimum width=1.45cm]
      (pred) at (5.8,1.2) {predicted\\rank $\widehat S$};
    \node[draw, rounded corners, align=center, fill=red!7, minimum width=1.65cm]
      (h) at (0,-1.2) {hidden $H$\\cache/policy state};
    \node[draw, rounded corners, align=center, fill=gray!10, minimum width=1.55cm]
      (y) at (3.0,-1.2) {policy cycles\\$(Y_a,Y_b)$};
    \node[draw, rounded corners, align=center, fill=orange!10, minimum width=1.45cm]
      (truth) at (5.8,-1.2) {true rank\\$S$};
    \draw[->, thick] (x) -- (z);
    \draw[->, thick] (z) -- (pred);
    \draw[->, thick] (x) -- (y);
    \draw[->, thick] (h) -- (y);
    \draw[->, thick] (y) -- (truth);
    \draw[dashed, red!70, thick] (h) --
      node[right, align=left] {not\\observed} (z);
  \end{tikzpicture}
  \caption{Partial observability in policy ranking. The true order depends
  on both the instruction window and hidden policy state, while every tested
  model can encode only the observed window.}
  \label{fig:partial-observability}
\end{figure}
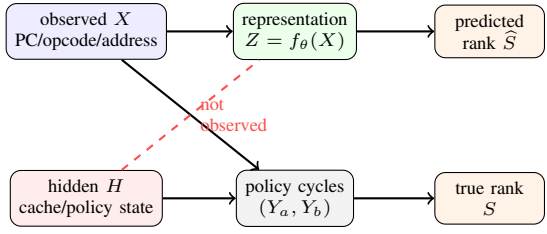

The predictors observe $X$ but not $H$.
Even with unlimited model capacity and training data, the best classifier
is bounded by the Bayes accuracy
\begin{equation}
  A^*_{ab}=\mathbb{E}_{X}
  \left[\max_{s\in\{-1,+1\}}P(S_{ab}=s\mid X)\right].
  \label{eq:bayes-accuracy}
\end{equation}
If different hidden states paired with the same observable window favor
opposite policies, $P(S=+1\mid X)$ remains near $1/2$ and
Equation~\eqref{eq:bayes-accuracy} remains near chance.
For any representation $Z$ computed only from $X$, the data-processing
inequality gives $I(S;Z)\leq I(S;X)$: a larger model extracts available
information more efficiently, but cannot create information about $H$ that
never entered its input.
Equivalently, when hidden state leaves $H(S\mid X)$ high, Fano's inequality
forces the Bayes error toward $1/2$, regardless of model size or inductive
bias.
Figure~\ref{fig:partial-observability} summarizes this separation between
model capacity and input observability.

The small gain over feature-free baselines across all four model families
is consistent with this low-information regime: the signal available in $X$
is real but sparse, and no increase in model expressiveness can recover
what was never observed.

\noindent\textit{In plain terms:} Which policy wins often depends on
things the instruction trace cannot tell you---whether a cache line is
currently hot, how full a prefetch queue is, what the replacement policy
evicted three thousand instructions ago.
No amount of architectural sophistication in the ML model can compensate
for missing inputs.
This is not a claim about the limits of ML in general; it is a claim about
what is knowable from the instruction stream specifically.

\subsection{Feature-Space Evidence}
\label{sec:tsne}

\begin{figure}[t]
  \centering
  \includegraphics[width=\columnwidth]{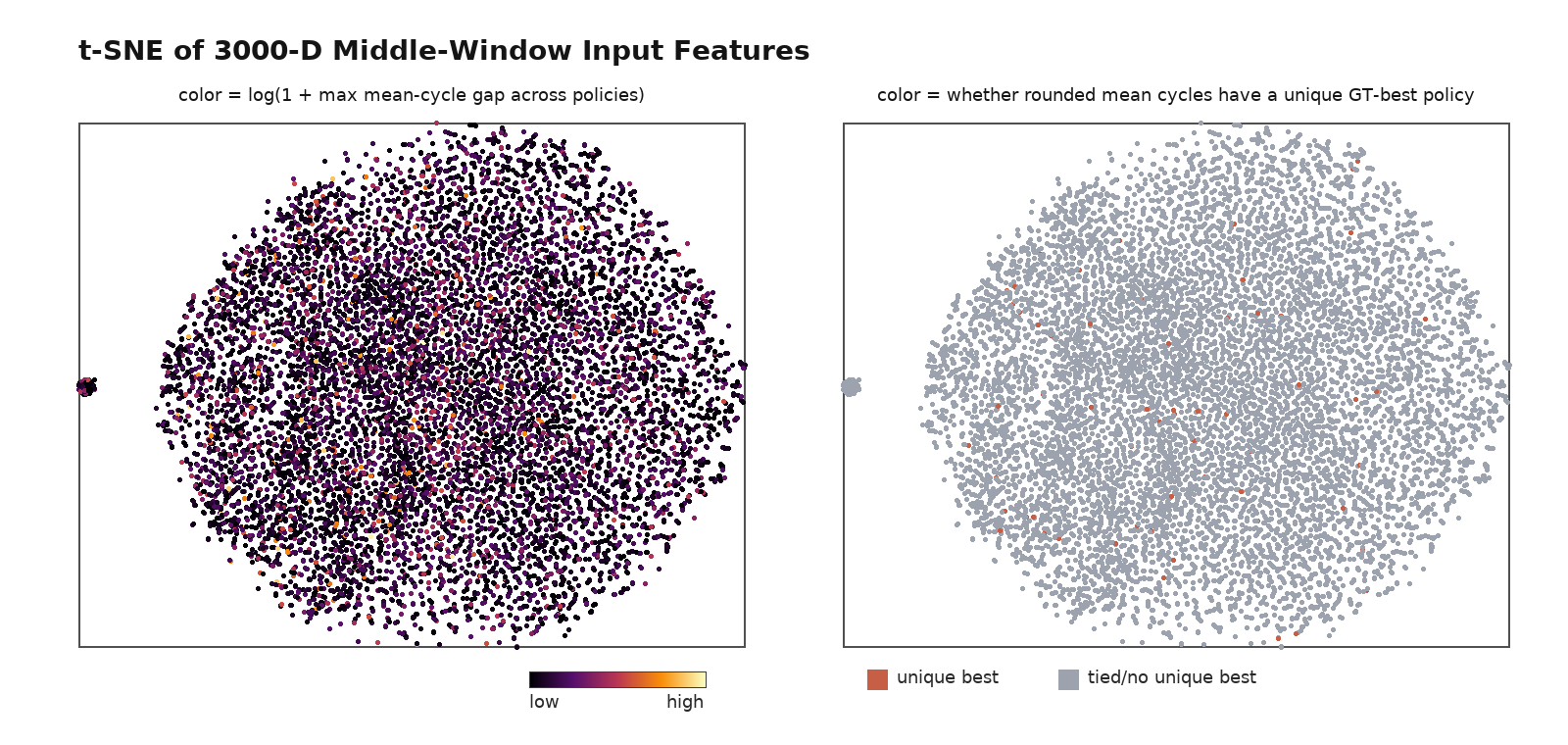}
  \caption{t-SNE of 3000-D \texttt{astar} middle-window inputs, colored by
  policy spread (left) and unique-best status (right). Neither forms clear
  clusters; this visualization is not proof of non-learnability.}
  \label{fig:tsne-margin}
\end{figure}

Figure~\ref{fig:tsne-margin} provides a qualitative view of observability
on sampled \texttt{astar} windows.
Each point flattens the features of the target instructions into a
3000-dimensional vector before t-SNE projection; color denotes policy spread
or unique-best status.
High-margin outcomes are sparse and do not form clean neighborhoods in the
projected input space.
The median policy spread is $0.14$ cycles and the 95th percentile is
$1.08$ cycles.

A projection can hide separability and t-SNE geometry is not globally
metric-preserving, so we do not infer impossibility from the plot.
Its role is narrower: it shows that the raw feature space does not contain
an obvious low-dimensional partition that the tested models somehow missed.
The quantitative evidence is the pairwise ranking result and its
baseline-adjusted confidence intervals---not the geometry of this
projection.

\noindent\textbf{In plain terms:}\emph{If instruction features reliably
separated ``policy A wins here'' windows from ``policy B wins here''
windows, we would see clusters in this visualization.
We do not.
This is consistent with the theoretical picture: the separating information
lives in the hidden state, not in the instruction features.}

\subsection{Why More Expressive Models Help Only Selectively}
\label{sec:expressiveness}

The four predictor families target different failure modes: the local CNN
tests short-range latency patterns, the recurrent model ordered context,
the summary model low-variance distributions, and the Transformer distant
interactions.
Greater expressiveness helps selectively: the best BP result is only
$2.09\%$ over majority and the best CIW result is $39.9\%$.

This pattern is expected under partial observability.
Model capacity helps when $X$ contains a nonlinear but recoverable signal,
as for \texttt{lbm} and several other benchmarks.
It cannot help where labels tie, margins fall below prediction noise, or
the winner is ambiguous without $H$.
The data indicate a mixture of regimes, not a universal absence of signal.
Because easy dominant cases can mask failure on reversals, a single global
accuracy number is unsafe for design selection.

\noindent\textit{In plain terms:} \emph{A bigger or smarter model only helps if
there is something in the instruction stream to learn from.
For benchmarks where one policy consistently wins during recognizable memory
phases, more expressive models find that pattern.
For most BP pairs, the winner changes based on state the model cannot see,
so expressiveness brings no benefit---the model is not underpowered, the
problem is simply not solvable from the available inputs.}

\section{Discussion and Implications}\label{sec:discussion}

These results do not argue against learned performance prediction; they
identify the regime boundary.
In the SP regime, ML reliably prunes and ranks configurations where resource
differences create large, systematic margins---the all-window result confirms
this directly.
In the BP regime, closely matched SOTA policies produce margins too small
and too dependent on hidden microarchitectural state for trace-based
predictors to rank reliably at instruction-window granularity.
The conclusion is conditional, not universal: benchmarks such as
\texttt{lbm} and \texttt{povray} contain repeated memory phases whose
observable patterns correlate with a consistent policy winner, showing that
signal exists where the instruction stream happens to encode the relevant
information.
These results apply to the six-field trace, nine BP
configurations, ChampSim simulator, and 29 SPEC CPU2006 benchmarks; they
do not establish that every possible state-aware learner must fail.

Prior work already demonstrates ML's value in tasks that do not require
per-window pairwise ranking, and our findings explain why those tasks
succeed where BP window-level ranking fails.
\emph{Coarse SP pruning}: our all-window SP accuracy of $77$--$89\%$
confirms that ML can reliably eliminate clearly inferior structural
configurations thus reducing a large sweep to a handful of finalists worth
simulating. Structural differences create margins large enough to
survive prediction error.
\emph{Single-configuration phase analysis}: SimNet~\cite{10.1145/3530891,10.1145/3656012}
demonstrates that ML can simulate a chosen processor design at
$100$--$1000\times$ cycle-level speed, identifying performance phases and
bottlenecks within that design; no cross-configuration ranking is required,
so the per-window ranking limit does not apply.
\emph{Aggregate-level DSE and attribution}: Concorde~\cite{10.1145/3695053.3731037} shows
that accurate absolute CPI prediction ($\sim$2\% error) enables both
benchmark-level configuration ranking and Shapley-style parameter
attribution at a scale infeasible for cycle-level simulation; our results
explain why this works---aggregate ranking benefits from the same
error-cancellation advantage that makes benchmark-level metrics look
stronger than per-window ones.
In each case the common thread is the same: ML succeeds when asked for
absolute performance estimates or coarse ordering over large margins, and
falls short only when asked to resolve narrow-margin per-window comparisons
where the outcome depends on hidden microarchitectural state.

Adding cache state, prefetch confidence, replacement metadata, and
outstanding-request information to the feature set would increase
observability and may improve BP ranking.
It also changes the deployment proposition: these features are expensive
to obtain without running a detailed timing model, are often specific to
one policy implementation, and risk leaking the very outcome the predictor
is meant to replace.
Thus, the relevant tradeoff is ranking information versus the cost and portability of obtaining it.
A practical middle ground is to learn compact state summaries from sparse
hardware-counter checkpoints, or to adopt a hierarchical strategy: use
architecture-independent ML to eliminate clearly inferior SP regions, then
apply direct simulation or state-aware models to the remaining BP
candidates.

\section{Related Work}\label{sec:related}
\textbf{Learned throughput and latency models.}
Ithemal predicts static basic-block throughput from assembly
instructions~\cite{mendis2019ithemal}. Similarly, GRANITE predicts basic-block throughput via graphical representation and processing~\cite{GRANITE}. SimNet predicts instruction latency
using static properties together with dynamic processor
state~\cite{10.1145/3530891, simnet_gpus, perfvec}, while NeuroScalar targets cycle-level prediction
from microarchitecture-independent traces~\cite{neuroscalar2026}. These works
establish that useful absolute performance quantities can be learned. Our
question is downstream and comparative: whether prediction error is small in
the differential direction needed to recover local reversals and order closely
matched policies.

\textbf{ML surrogates for simulation and DSE.}
Classical regression and interpolation methods model architecture parameters,
phases, performance, and power~\cite{li2009machine,zheng2016accurate,
ipek2006efficiently,joseph2006construction,joseph2006predictive,
lee2006accurate,lee2007illustrative,Simpoint_tool,simtrace}. Recently, ML provides these analysis~\cite{ML-power-model1,ML-power-model2,survey_arch_systems,NPC-DSE}. TAO uses reusable functional traces and
transfer learning across microarchitectures~\cite{10.1145/3656012}. Concorde
fuses analytical and ML components for fast CPU performance modeling and
attribution~\cite{10.1145/3695053.3731037}, and OneDSE uses workload-aware
Transformer prediction for metric estimation and
DSE~\cite{onedse2025}. These systems motivate our four-family evaluation, but
their objectives do not themselves establish strict pairwise rank
fidelity.

\textbf{Learning to choose policies.}
Prior work has studied ML-based runtime prefetcher selection and found that
selection quality depends on workload behavior and observable
features~\cite{alcorta2024characterizing}. Our setting differs in scale and
target: we align nine complete policy stacks and ask for all 36 strict pair
orders over held-out windows. In the structural space, we additionally isolate
the pairs with a clear architectural prior and their counter-intuitive
reversals. These tests expose label ties, small margins, and failures that
top-1 or aggregate IPC metrics can hide.

\textbf{Ranking versus regression.}
Prior predictor work primarily reports absolute error, speed, or average
CPI/IPC accuracy. DSE is comparative: a useful surrogate must preserve order
where candidates differ, identify the best candidate, and recognize when a
dominant hardware trend reverses. We show that regression accuracy and rank
fidelity can diverge, and that model accuracy must be compared against CIW
match, pair priors, tie rates, and the design space's margin distribution. Our
result complements prior predictors by identifying regimes in which their
standard success metrics are insufficient evidence for design selection.

\section{Conclusion}\label{sec:conc}

Fast and accurate cycle prediction does not imply reliable design ranking.
Across four predictor families evaluated under a common protocol, structural
configurations remain rankable in aggregate, but all four counter-intuitive-window
estimates fall below the $50\%$ random-ordering reference on the five
clear-prior pairs, showing that models recover dominant hardware trends
without detecting where those trends reverse.
In the behavioral-policy regime, $37.8\%$ of pair-windows are ground-truth
ties, most strict pairs carry margins of only a few cycles, and the best
selected model improves over a feature-free majority baseline by only $2.1\%$.
ML surrogates are genuinely useful for coarse structural pruning,
single-configuration phase analysis, and aggregate-level DSE---tasks where
large margins or absolute accuracy suffice---but should not be treated as
universal replacements for simulation when closely matched designs must be
ranked at instruction-window granularity, for deep microarchitecture
introspection.

\bibliographystyle{IEEEtranS}
\bibliography{references}

\end{document}